\documentclass{kluwer}    

\newdisplay{guess}{Conjecture}

\newcommand{\etal}{{et al.\ }}
\newcommand{\lta}{\stackrel{<}{\scriptstyle\sim}}
\newcommand{\gta}{\stackrel{>}{\scriptstyle\sim}}

\begin{document}                                                                                   
\begin{article}
\begin{opening}         
\title{Star Formation in Violent and Normal Evolutionary Phases} 
\author{Uta \surname{Fritze -- v. Alvensleben}}  
\runningauthor{U. Fritze -- v. Alvensleben}
\runningtitle{Violent and Normal Star Formationt}
\institute{Universit\"atssternwarte G\"ottingen}
\date{Dec. 1, 2001}

\begin{abstract}
Mergers of massive gas-rich galaxies trigger violent starbursts that -- over timescales of $> 100$ Myr and regions $> 10$ kpc -- form massive and compact star clusters comparable in mass and radii to Galactic globular clusters. The star formation efficiency is higher by 1 -- 2 orders of magnitude in these bursts than in undisturbed spirals, irregulars or even BCDs. We ask the question if star formation in these extreme regimes is just a scaled-up version of the normal star formation mode of if the formation of globular clusters reveals fundamentally different conditions.

\end{abstract}

\end{opening}           

\section{Modes of Star Formation}  
The basic question I'm going to address is if {\bf S}tar {\bf F}ormation ({\bf SF}) is a universal process just scaled up in intensity in the violent starbursts going on in merging gas-rich massive galaxies from what it is, at a much lower level, in normal spiral, dwarf, or low surface brightness galaxies, or if these vastly different sites feature different modes of SF.

\medskip\noindent
Normal SF in undisturbed galaxies produces stars, OB-associations, open clusters with masses up to ${\rm 10^3~M_{\odot}}$, and, sometimes, even a couple of Super Star Clusters (Larsen \& Richtler 1999). It does, however, not seem to produce {\bf G}lobular {\bf C}lusters (at least not in large numbers). Is GC formation a process different from that for open cluster/association/field star formation or is there {\bf one} universal process forming the entire continuum from associations through GCs? The answer may come from analyses of the molecular cloud structure and mass spectra or of the {\bf M}ass {\bf F}unction ({\bf MF}) of young star clusters, globular and open.  

\medskip\noindent
Violent SF occurs in interacting and merging gas-rich galaxies with the most powerful starbursts going on in Ultraluminous Infrared Galaxies with SF rates of several ${\rm 100~M_{\odot}yr^{-1}}$. Hydrodynamic modelling has shown that high gas pressure is built up in the inner parts of massive merging galaxies, observations have shown that sometimes enormous amounts of molecular gas at extremely high desities ${\rm > 1000~M_{\odot}pc^{-3}}$, comparable to the stellar densities in the cores of elliptical galaxies,  are assembled in regions of 1 -- few kpc extent. SF efficiencies are expected to be very high there. Indeed, the strong Balmer absorption lines and the still high luminosity of NGC 7252, a $\lta 1$ Gyr old merger remnant, under conservative assumptions concerning gas content G and luminosity of the progenitor spirals imply that over a scale of order 10 kpc the global starburst in this galaxy must have had a SF efficiency ${\rm \eta := \frac{\Delta S}{G}}$ of order 30 \% or more. This is at least a factor 10 higher than what is observed in spirals, irregulars, and BCDGs. It is high enough to allow for the formation of massive compact star clusters that may survive over a Hubble time, i.e. young GCs (Fritze -- v. A. \& Gerhard 1994, Fritze -- v. A. \& Burkert 1995).

Is the formation of GCs a criterium that discriminates between violent and normal SF regimes? What fraction of the Super Star Clusters observed in interacting galaxies are young GCs?

\section{Globular vs. Open Clusters}
In local undisturbed galaxies, there is a difference of $\gta 2$ orders of magnitude in mass between open clusters, most of which are younger than 1 -- 3 Gyr, and old GCs, their concentration parameters are different as well as their {\bf L}uminosity {\bf F}unctions ({\bf LF}). 
The high ambient pressure and high SF rates in interacting galaxies may or may not produce open clusters to higher masses than observed in non-interacting galaxies, tidal radii and concentration parameters are not accessible for clusters in mergers, leaving the focus on the LFs. The observed LFs of YSCs look like power-laws with slopes ${\rm m \sim -1.6~...~-1.8}$, similar to the slope of the Galactic or LMC open cluster LF. 

\smallskip\noindent
The old GC systems of the Milky Way, M31, and many E/S0 galaxies, in contrast, feature a roughly Gaussian LF with a turn-over ${\rm \langle M_V \rangle}$ sufficiently universal to be used for distance measurements and to constrain the Hubble constant. Their MF is log-normal with ${\rm \langle M_{GC} \rangle \sim 3 \cdot 10^5~M_{\odot}}$ (Ashman \& Zepf 1998). We do not know, however, the {\bf initial} MF of the Milky Way GC system at its formation. 

It used to be assumed that the initial MF of a GC system should reflect the power-law MF of the molecular clouds or molecular cloud cores out of which they form. Whereas in undisturbed galaxies the molecular cloud mass spectrum is a power-law, indeed, it has never been measured down to scales below an eventual turn-over in strongly interacting galaxies. The different molecular cloud density structure discussed above may cast doubts upon this assumption. Secular cluster destruction processes selectively destroying low-mass clusters were held responsible for the transformation of the assumed initial GC power-law MF into the log-normal one observed today (e.g. Ostriker \& Gnedin 1997). Taking internal gravitational forces into account, Vesperini (1997, 2000, 2001) found that the destruction of low-mass clusters by evaporation is balanced by the destruction of high-mass clusters through dynamical friction and that his models require considerable fine-tuning to transform an assumed initial power-law MF into a Gaussian, whereas an initial Gaussian GC MF evolves in a self-similar way despite destruction of about 50\% of the cluster population over a Hubble time, largely independent of his model parameters. 

Hence, the question as to the shapes of the molecular cloud (core) mass spectra and the initial cluster MF in interacting galaxies at present seems still open. While the resolution of molecular cloud (core) mass spectra even in the nearest interacting galaxies probably has to await ALMA, the MFs of YSCs can be analysed on HST data. 

\section{Mass Functions of Young Globular Clusters}
In ongoing starburst mergers OB-associations, open clusters, as well as GCs in significant numbers (Schweizer 2001) are forming. At an age of $\sim 1$ Gyr, most open clusters in the merger remnant NGC 7252 should have dissolved already and yet this object still features $\sim 500$ compact clusters with half-light radii typical of Milky Way GCs: a rich populations of young GCs (Miller \etal 1997). 

Meurer (1995) was the first to realise that age spread effects among clusters in ongoing or very recent starbursts may distort the LF with respect to the underlying MF. We confirmed and quantified this effect and tentatively derived for YSCs in the Antennae (NGC 4038/39) (Whitmore \& Schweizer 1995) a MF well compatibel with the log-normal MF of Milky Way GCs (Fritze -- v. A. 1998, 1999). Zhang \& Fall (1999), on the other hand, favor a power-law. Both approaches have their drawbacks. Since the data only give (V$-$I) (and (U$-$V) for a few clusters) we had to assume a uniform average reddening for all clusters. Zhang \etal used reddening-free color indices to cope with the apparently inhomogeneous dust. These have a non-monotoneous time evolution in a certain age range and they therefore excluded a number of clusters. If we exclude the age group of clusters that they excluded, we lose much of the age spread effect that accounts for the difference between the observed power-law LF and the Gaussian MF we find. \\
Hence, at present, the MF of YSCs is still controversal and it has only been investigated in one single galaxy. The basic shortcoming is that not enough passbands/colors are available to independently constrain individual YSC metallicities, dust reddening, and ages.

\section{Outlook}
Our ASTROVIRTEL project ``The Evolution and Environmental Dependence of Star Cluster Luminosity Functions'' (PI R. de Grijs) will provide a large sample of YSC systems with multi-color data combined from HST and ground-based telescopes. Homogeneous calibration and analysis  will 
allow to pin down the shape of the YSC MF and show how universal it is. This should take us an important step towards the answer to our question if violent SF is similar and only scaled up  with respect to ``normal'' SF in undisturbed galaxies or if it is intrinsically different. \\
In any case, it is clear that violent SF phases in galaxies are widely accompanied by the formation of populous cluster systems with many of their members surviving for many Gyr. Hence, the metallicity and age distributions of GC systems may contain valuable information about the (violent star) formation history of their parent galaxies. Again, multi-color data or individual GC spectroscopy are required to independently determine GC ages and metallicities. 
\acknowledgements
I gratefully acknowledge partial travel support from the EU and from the German Astronomische Gesellschaft.

\end{article}
\end{document}